\documentclass[prl,aps,reprint,superscriptaddress,floatfix]{revtex4-2}
\usepackage{amsmath}
\usepackage{amssymb}
\usepackage{bm}
\usepackage{graphicx}

\newcommand{\circled}[1]{\raisebox{.5pt}{\textcircled{\raisebox{-.9pt} {#1}}}}

\begin{document}
\title{Deterministic generation of skyrmions and antiskyrmions by electric current}

\begin{abstract}
Magnetic skyrmions are nanoscale spin whirlpools that promise breakthroughs in future spintronic applications. Controlled generation of magnetic skyrmions by electric current is crucial for this purpose. While previous studies have demonstrated this operation, the topological charge of the generated skyrmions is determined by the direction of the external magnetic fields, thus is fixed. Here, we report the current-induced skyrmions creation in a chiral magnet FeGe nanostructure by using the  \emph{in-situ} Lorentz transmission electron microscopy. We show that magnetic skyrmions or antiskyrmions can be both transferred from the magnetic helical ground state simply by controlling the direction of the current flow at zero magnetic field. The force analysis and symmetry consideration, backed up by micromagnetic simulations, well explain the experimental results, where magnetic skyrmions or antiskyrmions are created due to the edge instability of the helical state in the presence of spin transfer torque. The on-demand generation of skyrmions and control of their topology by electric current without the need of magnetic field will enable novel purely electric-controlled skyrmion devices.
\end{abstract}


\date{\today}

\author{Xuebing Zhao}
\email{These authors contributed equally to this work}
\affiliation{Laboratory of Advanced Materials, Collaborative Innovation Center of Chemistry for Energy Materials, Fudan University, Shanghai 200438, China}

\author{Jin Tang}
\email{These authors contributed equally to this work}
\affiliation{Anhui Key Laboratory of Condensed Matter Physics at Extreme Conditions, High Magnetic Field Laboratory, HFIPS, Anhui, Chinese Academy of Sciences, Hefei 230031, China}
\affiliation{University of Science and Technology of China, Hefei 230026, China}

\author{Ke Pei}
\affiliation{Laboratory of Advanced Materials, Collaborative Innovation Center of Chemistry for Energy Materials, Fudan University, Shanghai 200438, China}

\author{Weiwei Wang}
\affiliation{Institutes of Physical Science and Information Technology, Anhui University, Hefei 230601, China}

\author{Shi-Zeng Lin}
\email{szl@lanl.gov}
\affiliation{Theoretical Division, T-4 and CNLS, Los Alamos National Laboratory, Los Alamos, New Mexico 87545, USA}

\author{Haifeng Du}
\email{duhf@hmfl.ac.cn}
\affiliation{Anhui Key Laboratory of Condensed Matter Physics at Extreme Conditions, High Magnetic Field Laboratory, HFIPS, Anhui, Chinese Academy of Sciences, Hefei 230031, China}
\affiliation{University of Science and Technology of China, Hefei 230026, China}
\affiliation{Institutes of Physical Science and Information Technology, Anhui University, Hefei 230601, China}

\author{Mingliang Tian}
\affiliation{Anhui Key Laboratory of Condensed Matter Physics at Extreme Conditions, High Magnetic Field Laboratory, HFIPS, Anhui, Chinese Academy of Sciences, Hefei 230031, China}
\affiliation{University of Science and Technology of China, Hefei 230026, China}
\affiliation{Institutes of Physical Science and Information Technology, Anhui University, Hefei 230601, China}

\author{Renchao Che}
\email{rcche@fudan.edu.cn}
\affiliation{Laboratory of Advanced Materials, Collaborative Innovation Center of Chemistry for Energy Materials, Fudan University, Shanghai 200438, China}

\maketitle

Magnetic skyrmion is a stable spin texture with nanometer scale size and nontrivial topology
~\cite{Rossler_Spontaneous_2006, Muhlbauer_Skyrmion_2009, Yu_Realspace_2010, 
Qin_Noncollinear_2020, Tang_Magnetic_2021, Wei_Dzyaloshinsky_2021},
and is considered as information carrier of binary digits or logical element in a number of futuristic spintronic devices
including racetrack memories~\cite{Fert_Skyrmions_2013}, logic gates~\cite{Zhang_Magnetic_2015},
probabilistic computing~\cite{Pinna_Skyrmion_2018}, and even neuromorphic devices~\cite{Song_Skyrmionbased_2020}.
The nontrivial topology of a skyrmion is described by a quantized topological charge $Q$ according to
$Q=(1/4\pi)\int{dr^2\mathbf{n}\cdot(\partial_x\mathbf{n}\times\partial_y\mathbf{n})}$ with a unit vector $\mathbf{n}$
representing the direction of magnetic moments~\cite{Nagaosa_Topological_2013}.
Like the fundamental particles in nature, a skyrmion ($Q = +1$) has its corresponding antiparticle or antiskyrmion
with opposite topological charge ($Q = -1$)~\cite{Nayak_Magnetic_2017, Bera_Theory_2019, Peng_Controlled_2020}.
The skyrmion and antiskyrmion are linked by time reversal operation, and therefore the application of external
magnetic field determines if the skyrmions or antiskyrmions are stabilized in the system. The previous study on
the electrical generation of skyrmions requires external magnetic fields~\cite{Finizio_Deterministic_2019,
Woo_Deterministic_2018, Brock_Currentinduced_2020, Akhtar_CurrentInduced_2019, Wei_Current_2021, Jiang_Blowing_2015,
Lemesh_CurrentInduced_2018, Legrand_RoomTemperature_2017, Wang_Thermal_2020, Je_Targeted_2021},
which in turn determine the topological charge of the created skyrmions. Furthermore, 
the static magnetic field required to stabilize magnetic skyrmion complicates the device design
and increases the energy consumption. To generate skyrmions with a controlled topological charge electrically,
it is important to stabilize skyrmions without the need of external magnetic fields. However,
most systems require magnetic field to stabilize skyrmions with only few
exceptions~\cite{Gallagher_Robust_2017, Karube_Skyrmion_2017, Yu_Aggregation_2018}.

Here, we report on controlled generation of skyrmions or antiskyrmions by tuning the direction of electric current 
in FeGe at zero magnetic field. We reveal the topological charge in term of a half skyrmion localized at the ends of 
the spin helix. A force induced by spin transfer torque (STT) from electric current acts on the half skyrmion, or meron, 
and tends to detach the half skyrmion from the helix~\cite{Zhang_Roles_2004, Everschor_Rotating_2012}. 
For current above a threshold value, the half skyrmion is split off the helix and is relaxed into a full skyrmion 
or antiskyrmion determined by the direction of the force with respect to the orientation of the spin helix 
and hence the direction of the current flow.

We choose the chiral magnet FeGe to demonstrate the creation of magnetic skyrmions and antiskyrmions
~\cite{Tang_Magnetic_2021, Yu_roomtemperature_2011}. In chiral magnets, magnetic skyrmions are commonly 
characterized by three parameters: the core polarity, $p = [1, -1]$ (up or down of the magnetic moments), 
vorticity $W = [1, -1]$ (vortex or antivortex) and in-plane circular magnetization surrounding the core, 
i.e., helicity c. The topological charge is $Q = pW$. The Dzyaloshinsky-Moriya Interaction (DMI) in FeGe 
selects $W = 1$ and locks $c = [\pi/2, -\pi/2]$ (clockwise or counterclockwise) with $p$ such that $cp=-\pi/2$.
The polarity $p$ is always antiparallel to the direction of the external magnetic field, and therefore 
both the topological charge and skyrmion helicity are determined by magnetic field direction.
At zero field, the skyrmion ($Q = +1$) and antiskyrmion ($Q=-1$) are degenerate in energy.
We use the Lorentz TEM to detect the parameters $c$ and $p$. In our experiments with a negative defocus value, 
white dots correspond to skyrmions, while antiskyrmions are observed as black dots.

\begin{figure}[t]
\begin{center}
\includegraphics[width=0.45\textwidth]{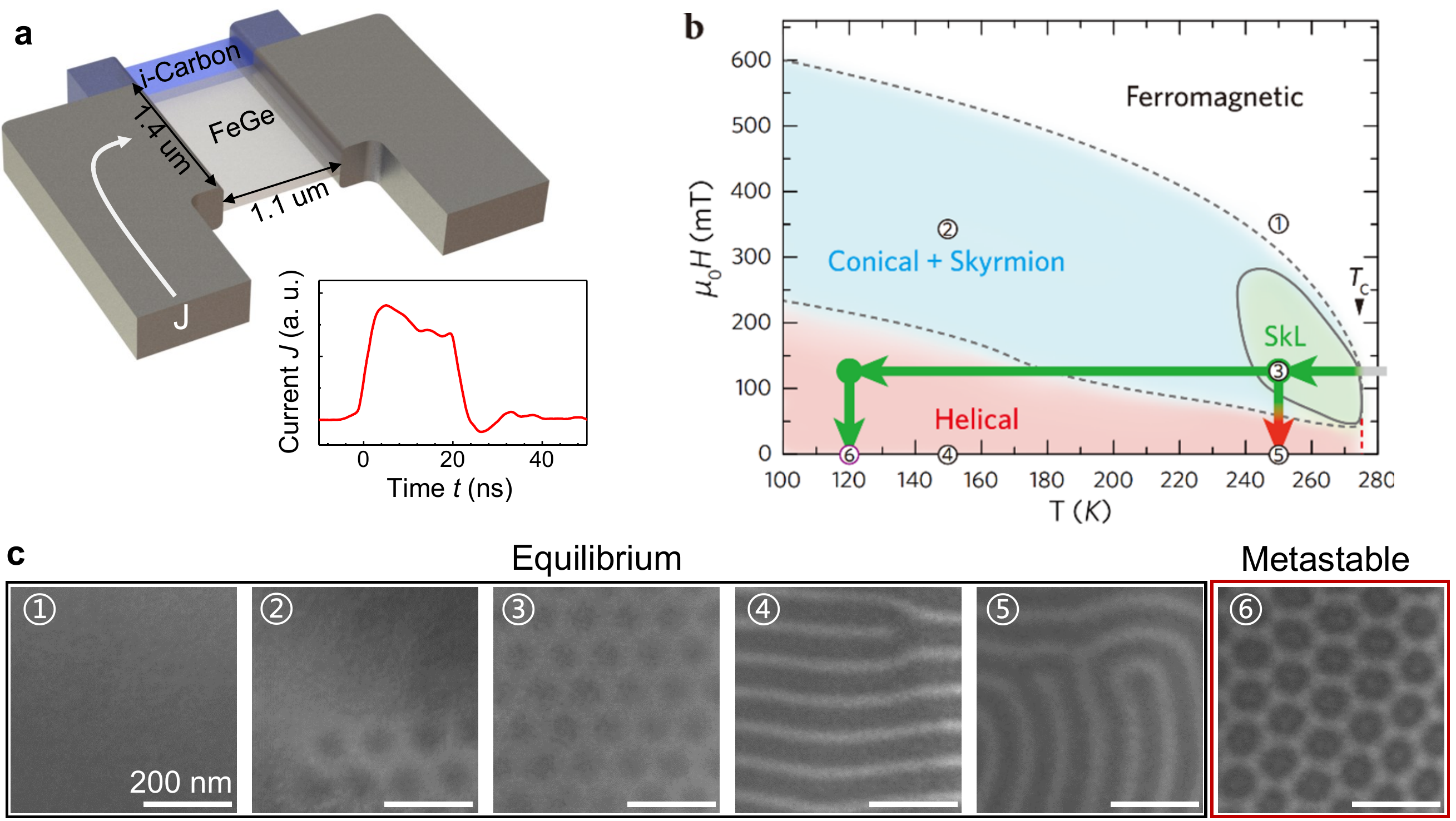}
\caption{Experimental set-up and magnetic phase diagram of an FeGe thin plate. (a), Schematic diagram of a 
microdevice composed of a $\sim$ 100-nm-thick thin plate for in-situ Lorentz TEM observation and thicker parts 
for electric contacts. Inset shows a representative line profile of a single current pulse with duration of 20 ns. 
(b), B-T phase diagram of the plate and Lorentz images of representative magnetic configurations of \circled{1}, 
ferromagnetic state, \circled{2}, mixed conical and isolated antiskyrmions states, \circled{3}, antiskyrmion lattice (aSkL), 
\circled{4} and \circled{5} helical sate, \circled{6}, metastable antiskyrmion lattice that is obtained by field cooling procedure. }
\label{fig1}
\end{center}
\end{figure}

Fig. 1a shows the schematics of the microdevice used in this experiment. The device is composed of two electrodes 
and a thin region with a thickness of $\sim$ 100 nm for \emph{in-situ} Lorentz TEM observation~\cite{Tang_Lorentz_2019}. 
The current pluses were applied through the two electrodes and set to have a pulse width of 20 ns. 
The typical magnetic field-temperature ($B-T$) phase diagram of the thin FeGe plate is shown in Figs. 1b and 1c. 
In the low $B$ region, the spin helix is the ground state~\cite{Yu_roomtemperature_2011}. Upon increasing positive $B$, 
the helix transforms into isolated magnetic antiskyrmions (black dots under the negative defocused conditions) in the conical 
background at low temperature for $T < 240$ K. For $T > 240$ K, a thermodynamically stable antiskyrmion lattice (aSkL) emerges. 
The spins are fully polarized in the high $B$ region. The transitions between these states are of first order~\cite{Huang_Melting_2020}, 
and shows a strong hysteresis. This $B-T$ phase diagram is generic for the nanostructures of chiral magnets~\cite{Nagaosa_Topological_2013}.

Owing to the nature of the first order phase transition~\cite{Huang_Melting_2020}, metastable magnetic states can be created 
by controlling their histories of sweeping $B$ and $T$~\cite{Zhao_Direct_2016}. Taking the point \circled{6} in the phase diagram 
for example (Fig.~1b), the corresponding ground states are spin helix. However, if one follows the path 
\circled{3}-\circled{4}-\circled{6}, a metastable zero-field SkL can be obtained (Fig.~1c). 
The $Q$ of such zero-field metastable antiskyrmion lattice is the same as that of the parent state in 
the point \circled{3}, and hence is determined by the direction of $B$. The metastability nature of antiskyrmions 
at $B \sim 0$ mT is crucial for the antiskyrmion generation by current with a reversible $Q$. 
Otherwise, the electrically generated skyrmions from the spin helix disappear quickly after their creation 
at high temperature $T > 240$ K.

\begin{figure}[t]
\begin{center}
\includegraphics[width=0.45\textwidth]{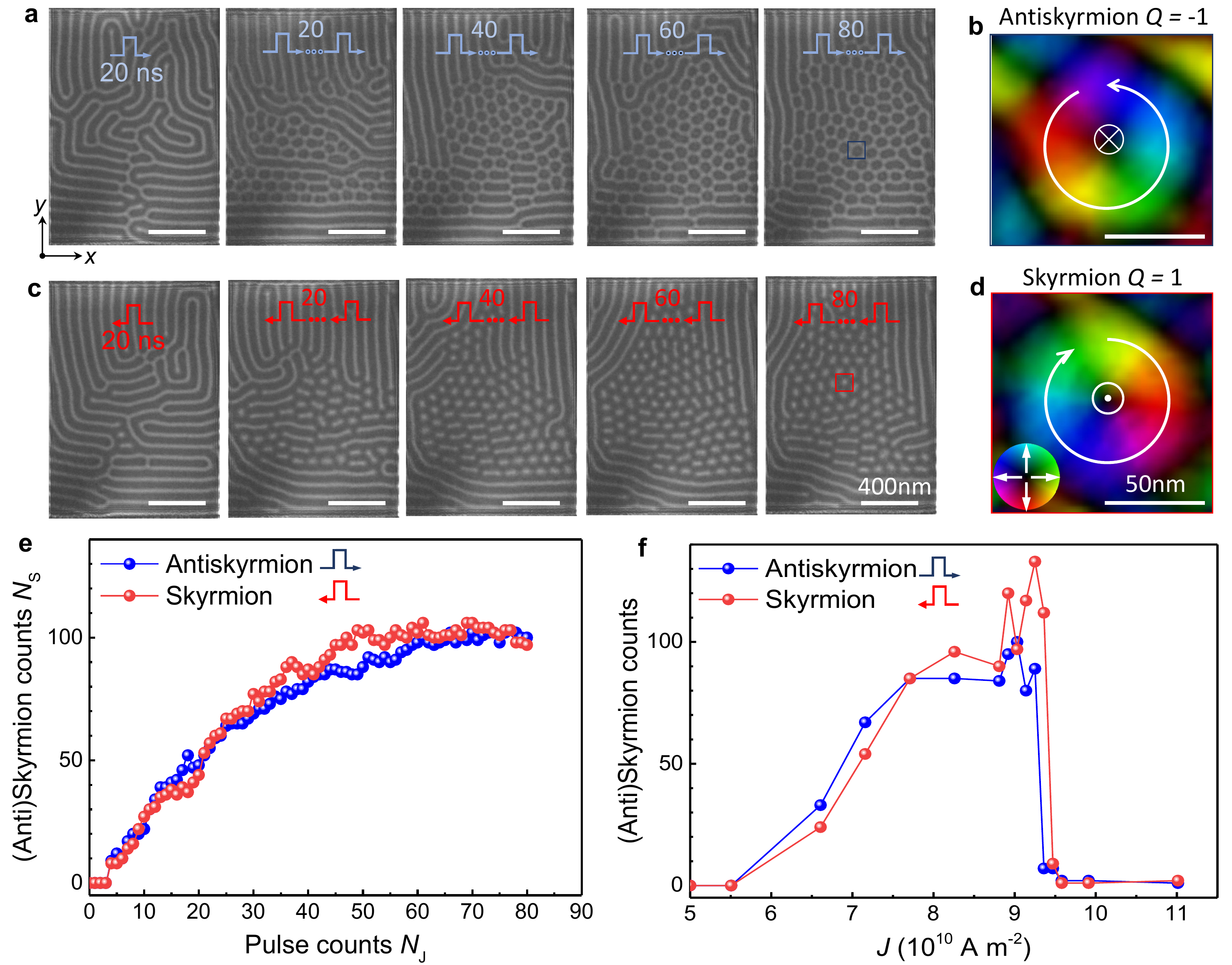}
\caption{Skyrmions and antiskyrmions generation using currents at zero magnetic field and the temperature 
of $T \sim 95$ K. (a), Snapshots of Fresnel images taken at initial state, after 20 current pulses, 40 current pulses, 
60 current pulses, and 80 current pulses, respectively. The current with the density $J \sim 9.0 \times 10^{10}$ A m$^{-2}$ 
is applied along the $x$ axis. (b), In-plane magnetization mapping of an antiskyrmion with $Q = –1$ in the blue region of (a), 
the helicity is determined as $c=\pi/2$. (c), Snapshots of Fresnel images taken at initial state, after 20 current pulses, 
40 current pulses, 60 current pulses, and 80 current pulses, respectively. The current density is the same to that in (a) and 
the direction is reversed to be along the $-x$ axis. (d), In-plane magnetization mapping of a skyrmion with $Q = 1$ in the red 
region of (c), the helicity is determined as $c=-\pi/2$. (e), Count of created magnetic skyrmions (blue dots) or antiskyrmions 
(red dots) as a function of pulse counts. (f), Current density of maximum (anti)skyrmion counts created by pulsed current 
from initial helical domains.
}
\label{fig2}
\end{center}
\end{figure}

The experimental results of the skyrmions or antiskyrmions creation by a pulsed current at zero-field are summarized 
in Fig.~2. We set the pulse width of the current to be 20 ns and the frequency to be 1 Hz on the FeGe microdevice 
(Fig.~1a and supplemental Fig.~S1). The initial magnetic state is the spin helix (Fig.~2a), which approximately propagates 
in horizontal (vertical) direction in the upper (lower) part of the sample. In the central region of the sample, 
a half-skyrmion structure forms at the end of the helix and will play a key role to create skyrmions~\cite{Ezawa_Compact_2011}, 
which will be discussed below. When a series of pulsed currents with the density of $J \sim 9.0 \times 10^{10}$ A m$^{-2}$ 
is applied along the $x$ direction the spin helices shake collectively accompanied with the creation of spin textures 
(Fig. 2a and Supplementary Video 1). These spin textures are identified as antiskyrmions with $Q = –1$ from Lorentz TEM images 
(Fig.~2b). The count of created antiskyrmions increases with the pulse number until $\sim 60$ pulses, after which the skyrmion 
number saturates (Fig. 2e). Remarkably, skyrmions with $Q = 1$ can be generated simply by reserving the direction of current 
pulses along the $-x$ axis, as shown in Fig.~2c and Supplementary Video 1, where both the helicity and polarity of the spin 
textures are reserved (Fig.~2d).

Effects of current density on the maximum count of (anti)skyrmions created by pulsed current at zero-field are then explored 
(Fig.~2f and Supplemental Fig.~S2). The generation of skyrmions requires a threshold current 
$J_{c1} \sim 5.5 \times 10^{10}$ A m$^{-2}$, below which the initial spin helices have a weak dynamic response to the pulsed currents 
(Supplemental Fig.~S3). The maximum count of (anti)skyrmions increases as current density $J$ increases (Supplemental Figs.~S4 and S5). 
However, the maximum count of (anti)skyrmions decreases to $\sim 0$ once the current density $J$ is above an upper 
threshold current density $J_{c1} \sim 9.4 \times 10^{10}$ A m$^{-2}$ (Supplemental Fig.~S6), which is attributed to the 
destabilization of skyrmions as a consequence of the significant temperature rise due to the large Joule heating 
effect~\cite{Zhao_Thermal_2018}.

\begin{figure}[t]
\begin{center}
\includegraphics[width=0.45\textwidth]{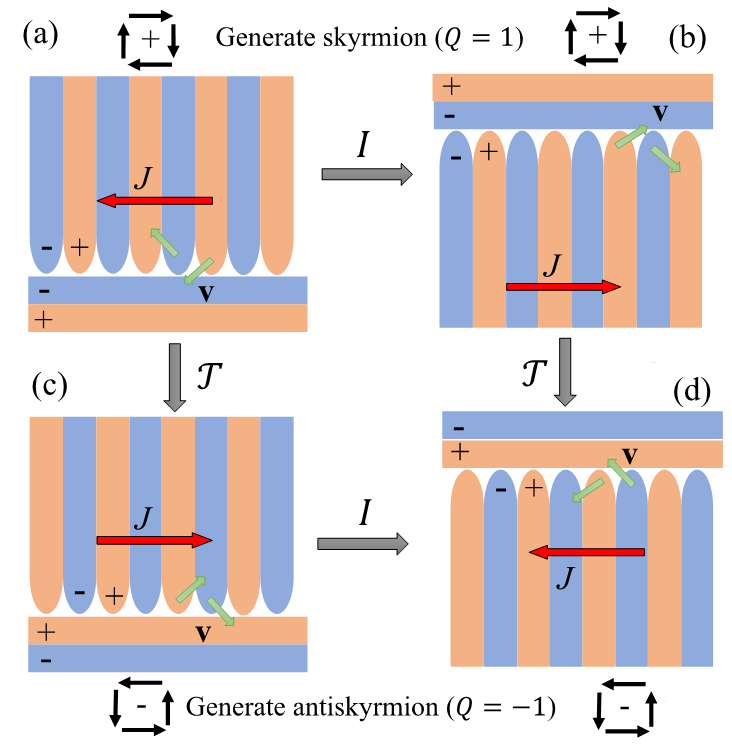}
\caption{Skyrmion or antiskyrmions generation rule based on force and symmetry analysis. Four types of helix stripes 
motivated by experiments are shown for simple demonstration of the underlying physics. Helix stripes with different polarity, 
$p = –1$ and 1 are denoted by blue (-) and orange (+), respectively. The relative motion of the merons/antimemons 
in the presence of a current (red arrow) is represented by green arrows.  Starting from (a), other cases in (b), (c) and (d) 
can be obtained by applying time reversal $\mathcal{T}$ and/or combined rotation of spin and space by 180 degrees 
$\mathcal{I}$ to (a). The topological charge of the generated skyrmions/antiskyrmions is also shown.
}
\label{fig3}
\end{center}
\end{figure}

To understand the skyrmion creation from the spin helix by current, we analyze the force on the spin helix based 
on the Thiele equation~\cite{Thiele_SteadyState_1973}. For a straight helix, the spins are coplanar inside the helix, 
and therefore the skyrmion topological charge vanishes. However, at the end points of the helix, there exist a half skyrmion 
(or meron) with a topological charge $Q_m=\pm1/2$ and the sign depends on the out-of-plane component of the spin 
(Figs.~2 and 4)~\cite{Ezawa_Compact_2011}. While the main body of the straight helix is inert to the electric current, 
the meron at its end couples to the current through STT, and is responsible for the skyrmion generation. 
By treating the meron as a rigid particle, we can obtain the equation of meron motion using Thiele’s collective 
coordinate approximation~\cite{Everschor_Rotating_2012, Thiele_SteadyState_1973, Ezawa_Compact_2011, Lin_Edge_2016, 
Iwasaki_Universal_2013}:
\begin{equation}\label{eq1}
  Q_m\mathbf{z}\times\left(\mathbf{v}_\mathbf{s}-\mathbf{v}\right)+\eta\left(\beta\mathbf{v}_s-\alpha\mathbf{v}\right)+\mathbf{F}=0
\end{equation}
where $\mathbf{z}$ is a unit vector perpendicular to the film, $\mathbf{v}$ is the skyrmion velocity and the form factor of 
a meron is $\eta_\mu=(1/4\pi)\int{dr^2\left(\partial_\mu n\right)^2}$ which we assume $\eta=\eta_x=\eta_y$ for simplicity. 
$\alpha$ is the Gilbert damping constant, $\beta$ is the coefficient of the non-adiabatic spin transfer torque, 
and $\mathbf{F}$ denotes all the other forces including inter-meron interaction, meron-defect interaction 
and the line tension between meron and the helix. $\mathbf{v}_s$ is the charge carrier velocity, which is parallel 
(antiparallel) to current $\mathbf{J}$ for hole (electron) carriers. When $\mathbf{F}$ is neglected, 
a meron has a velocity component antiparallel to current 
$\mathbf{v}_\parallel=\left(\frac{\beta}{\alpha}+\frac{\left(\alpha-\beta\right)Q_m^2}{\alpha^3\eta^2+\alpha Q_m^2}\right)\mathbf{v}_s$ 
that is independent on the sign of $Q_m$, and a velocity component perpendicular to the current 
$\mathbf{v}_\bot=\frac{\left(\alpha-\beta\right)\eta Q_m}{\alpha^2\eta^2+Q_m^2}\mathbf{z}\times\mathbf{v}_s$ 
that depends on the sign of $Q_m$. Recent experiments on the current-induced skyrmions motion in FeGe show that 
the main carriers are holes~\cite{Dong_Shooting_}.

\begin{figure}[t]
\begin{center}
\includegraphics[width=0.45\textwidth]{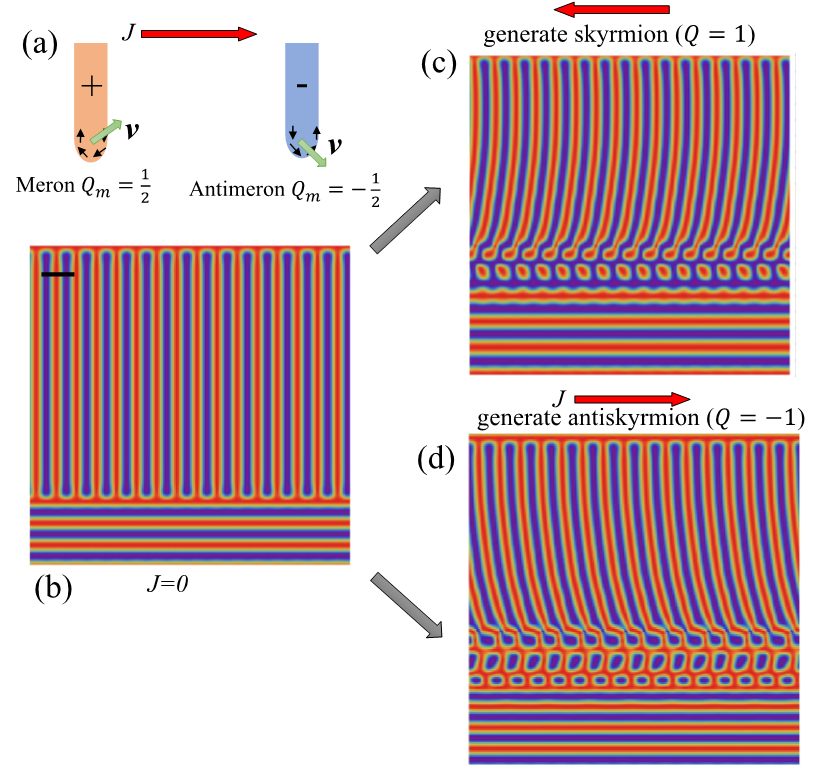}
\caption{Simulated results of skyrmion or antiskyrmions generation. (a), A half meron or antimeron with $Q_m=\pm1/2$ at 
the ends of the helix stripe. The sign of the meron charge depends on the polarity of the helix stripe. 
The direction of motion for meron/antimeron under an electric current $\mathbf{J}$ is shown by arrows. 
(b), Initial magnetic configurations of spin helix. Starting from the helix stripe in (b), 
either skyrmions in (c) or antiskyrmion in (d) are created depending on the direction of current. 
The length scale of the images is denoted by the bar in (b) with length equal to $20 J_{ex}/D$. 
See Supplementary Videos 2 and 3 for animation.
}
\label{fig4}
\end{center}
\end{figure}

For a weak current, a new stationary state for the meron is reached by the line tension through deforming the helix. 
For a large current, the line tension is not sufficient to balance the force due to current, and as a consequence, 
the meron is detached from the spin helix. The meron is an unstable configuration and relaxes into a full skyrmion 
in the end. This picture was invoked to explain the skyrmion generation by blowing the stripe domain through a 
constriction~\cite{Jiang_Blowing_2015, Lin_Edge_2016}.

The dependence of topological charge of the created skyrmion on the direction of current and the helix configuration 
can be understood in terms of the aforementioned physical picture. For a helix configuration in Fig. 3a as observed 
in experiment, the helix stripe is stretched or compressed depends on the $Q_m$ and hence on the polarity of the helix stripe. 
When a current is directed to the left, the stripes with +1 (up) polarity is stretched while the – (down) polarity is compressed. 
The merons at the end of the stripe then is split off when they are maximally stretched, while these compressed remain attached 
to the stripe. As a consequence, the generated skyrmions have topological charge $Q = 1$. Similarly, we can determine the $Q$ 
in the other helix configurations based on symmetry transformation by noting that the time-reversal symmetry breaking 
due to the damping (Gilbert damping and the non-adiabatic spin transfer torque contribution) is very weak for FeGe. 
Moreover, the DMI term responsible for the stabilization of skyrmions in FeGe, $D\mathbf{n}\cdot(\nabla\times\mathbf{n})$, 
is invariant under the combined rotation of spin and space along the axis perpendicular to the film (z axis) by 180 degrees. 
For a given meron charge $Q_m$, both $v$ and $J$ changes sign under the time reversal $\mathcal{T}$ and the combined 
rotation of spin and space $\mathcal{I}$. Under $\mathcal{T}: J\rightarrow-J$, $n\rightarrow-n$, spatial 
coordinate $r\rightarrow r$, we transform the case in Fig. 3a to Fig. 3c. The skyrmion topological charge changes 
sign under $\mathcal{T}$, therefore the generated skyrmion has the opposite sign in Fig.~3c compared to Fig.~3a. 
Under $\mathcal{I}$: $J\rightarrow-J, n_{x,y}\rightarrow-n_{x,y}, n_z\rightarrow n_z$, in-plane spatial coordinate 
$r\rightarrow-r$, we transform the Fig.~3a to Fig.~3b. The $Q$ is invariant under $\mathcal{I}$, and the $Q$ in Fig.~3a 
to Fig.~3b is the same. Under $\mathcal{T}\times \mathcal{I}$, we have situation in Fig.~3d.

Going beyond the qualitative analysis, we also perform the micromagnetic simulation by solving Landau-Lifshitz-Gilbert (LLG) 
equation in two dimensions. Motivated by the experiments, we choose the helix configuration in Fig.~4b and assume that the 
main body of the helix is strongly pinned by impurities in the systems. Therefore, we apply an electric current only 
in the lower part of the system to create magnetic skyrmions. The $Q$ of created skyrmions depends on the direction 
of the current (Fig.~4), and is consistent with simple picture in Fig.~3 and also the experiments in Fig.~2. 
The threshold current for skyrmion generation at zero temperature is of the order of $10^{12}$ A/m$^2$. 
At higher $T$ close to the Curie temperature, the helix stripe is less stiff, therefore the threshold current 
is expected to be much smaller (Fig. 2f). On-demand generation of skyrmions by electric currents is a crucial 
for skyrmion device applications. Néel-type skyrmions generations have been demonstrated previously in magnetic 
heterostructures with the requirement of external magnetic 
field~\cite{Jiang_Blowing_2015, Lemesh_CurrentInduced_2018, Legrand_RoomTemperature_2017, Wang_Thermal_2020, Je_Targeted_2021}, 
where the topological charge is selected by the direction of external magnetic field. 
The skyrmion generation demonstrated here is the first experiment of deterministic creation of Bloch-type skyrmion 
without the need of magnetic field, where the topological charge can be reversed by changing the direction of the electric current. 
The skyrmions are created out of the spin helix, which is ubiquitous and robust in chiral magnets. 
Combined with electric detection of skyrmions scheme~\cite{Wang_Electrical_2019}, e.g., 
the anomalous Hall effect~\cite{Maccariello_Electrical_2018}, the controlled electrical generation of skyrmions 
in FeGe nanostructure point to a promising direction for advanced skyrmionic device concept, such as the racetrack memory.

This work was supported by the National Key R\&D Program of China, Grant No. 2017YFA0303201 and 2018YFA0209102; 
the National Natural Science Foundation of China, Grant No. 51725101, 11727807, 51672050, 61790581, 11804343, and 12174396; 
the Strategic Priority Research Program of Chinese Academy of Sciences, Grant No. XDB33030100; 
and the Equipment Development Project of Chinese Academy of Sciences, Grant No. YJKYYQ20180012. 
The work at LANL was carried out under the auspices of the U.S. DOE NNSA under contract No. 89233218CNA000001 
through the LDRD Program, and was supported by the Center for Nonlinear Studies at LANL.

\bibliographystyle{apsrev4-1}
\bibliography{skx_creation}

\end{document}